\documentclass[pra,aps,showpacs,twocolumn]{revtex4}

\usepackage{epsf}
\usepackage{amsmath}
\usepackage[dvips]{color}
\newcommand {\be}  {\begin{equation}}
\newcommand {\ee}  {\end{equation}}
\newcommand {\bea} {\begin{eqnarray}}
\newcommand {\eea} {\end{eqnarray}}
\newcommand {\dd}  {\mathrm{d}}
\newcommand {\sn}  {\mathrm{sn}}

\begin{document}


\title{Nonlinear Band Structure in Bose Einstein Condensates: The Nonlinear Schr\"odinger Equation with a Kronig-Penney Potential}
\author{B. T.  Seaman, L. D. Carr and M. J. Holland}
\affiliation{JILA, National Institute of Standards and Technology and Department of Physics,
\\University of Colorado, Boulder, CO  80309-0440}
\date{\today}


\begin{abstract}
All Bloch states of the mean field of a Bose-Einstein condensate in the presence of a one dimensional lattice of impurities are presented in closed analytic form.  The band structure is investigated by analyzing the stationary states of the nonlinear Schr\"odinger, or Gross-Pitaevskii, equation for both repulsive and attractive condensates.  The appearance of swallowtails in the bands is examined and interpreted in terms of the condensates superfluid properties.  The nonlinear stability properties of the Bloch states are described and the stable regions of the bands and swallowtails are mapped out.  We find that the Kronig-Penney potential has the same properties as a sinusoidal potential; Bose-Einstein condensates are trapped in sinusoidal optical lattices.  The Kronig-Penney potential has the advantage of being analytically tractable, unlike the sinusoidal potential, and, therefore, serves as a good model for experimental phenomena.
\end{abstract}


\maketitle

\section{Introduction}
\label{sec:Introduction}

Periodic potentials are ubiquitous in physics, appearing in
electron transport in metals~\cite{Altmann1970}, Josephson
junction arrays~\cite{Usmanov2004}, nonlinear photonic crystals
and waveguide arrays~\cite{Christodoulides2003}, and Bose-Einstein
condensates (BEC's)~\cite{Greiner2001}.  With the realization of
BEC's of alkali atoms in a sinusoidal optical lattice, there has
been an explosion in studies of BEC's in periodic potentials, both
experimentally and
theoretically~\cite{Greiner2002,Fallani2003,Eiermann2003,Esslinger2003,Eiermann2004,Fallani2004}.
BEC's in periodic potentials have been used to study phase
coherence of atom lasers~\cite{Anderson1998,Hagley1999} and
matter-wave diffraction~\cite{Ovchinnikov1999}.  Therefore, in the
context of the BEC, the study of periodic potentials provides an
excellent connection between condensed matter physics and atomic
physics.  In contrast to other physical contexts, the lattice
geometry and strength, as well as the interatomic
interactions~\cite{Roberts1998,Inouye1998}, are all tunable
parameters for the BEC.  We show that the mean field Bloch states
of a BEC in a Kronig-Penney potential, i.e., a lattice of delta
functions, exhibits the same band structure and stability
properties as the experimental case of a sinusoidal potential.
Unlike in the case of the sinusoidal potential, Bloch state
solutions to the Kronig-Penney potential can be described by
straightforward analytic expressions.  The Kronig-Penney potential
has distinct advantages as a model of BEC's trapped in periodic
potentials.

Specifically, we consider the steady state response of the mean
field of a BEC to a Kronig-Penney potential using the Bloch
ansatz.  The mean field is modeled by the nonlinear Schr\"odinger
equation (NLS), which appears in numerous areas of physics; in the
context of the BEC, it is often called the Gross-Pitaevskii
equation~\cite{Gross1961,Pitaevskii1961}.  We have previously
obtained the full set of stationary solutions for a single delta
function
analytically~\cite{Seaman2004,Bogdan1997,Astrakharchik2004,Radouani2004}.
Using these results, we are able to rigorously describe the band
structure.  We find that above a critical nonlinearity,
swallowtails, or loops, form in the
bands~\cite{Wu2002,Machholm2003}.  These swallowtails are related
to superfluid properties of the BEC~\cite{Mueller2002}, as we
shall explain.  Stability properties are studied numerically by
time evolution of perturbed stationary state solutions to the NLS.
It is found that stable, as well as unstable, regimes exist for both
repulsive and attractive BEC's.  We also elucidate the linear
Schr\"odinger and discrete nonlinear Schr\"odinger limits of our
model.  The latter limit is related to the superfluid phase of the
Bose-Hubbard model~\cite{Rey2003}.

Experimentally, in order to create a BEC in a lattice, alkali
atoms are first cooled to a quantum degenerate regime by laser
cooling and evaporation in harmonic electromagnetic
traps~\cite{Anderson1995,Davis1995}.  A sinusoidal optical lattice
is then created by the interference of two counter-propagating
laser beams, which creates an effective sinusoidal potential
proportional to the intensity of the beams~\cite{Greiner2002}. The
potential is due to the AC Stark shift induced by the dipole
interaction with the electromagnetic field on the atoms' center of
mass~\cite{Meystre1999}.  For large detuning of the optical field
from the atomic transition, dissipative processes, such as
spontaneous emission, can be minimized and the potential becomes
conservative.  Nonzero quasimomentum can be examined by slightly
detuning the two lasers by a frequency $\delta
\nu$~\cite{Fallani2004}.  The resulting interference pattern is
then a traveling wave moving at the velocity, $v =
(\lambda/2)\delta \nu$, where $\lambda$ is the wavelength of the
first beam.  This produces a system with quasimomentum
$q=mv/\hbar$, where $m$ is the atomic mass.  After a given
evolution time, the traps are switched off.  The BEC is allowed to
expand and the density is then imaged.

The sinusoidal optical lattice potential is composed of a single
Fourier component.  If more counter-propagating laser beams of
different frequencies are added, more Fourier components are
introduced, and the potential becomes a lattice of well separated
peaks.  In the limit that the width of the these peaks becomes
much smaller than the healing length of the BEC, the potential
effectively becomes a Kronig-Penney lattice.

We proceed to highlight a few of the many experiments on BEC's
trapped in lattice potentials.  In a work by Anderson and
Kasevich~\cite{Anderson1998}, the tunneling of a BEC between
sites of an optical lattice aligned with the gravitational field
was examined.  This produced the matter wave analog of the AC
Josephson effect, as well as a new kind of atom laser.  Greiner
{\it et al.}~\cite{Greiner2002} observed a quantum phase
transition in a 3D lattice from a superfluid to a Mott-insulator.
BEC's trapped in lattices have been proposed as one possible
realization of a quantum computer and many studies have been
performed in this direction, see for example~\cite{Cirac2004} and references therein.  An
optical lattice was used to renormalize the effective mass to
create a Tonks-Girardeau
gas~\cite{Girardeau1960,Paredes2004,Kinoshita2004}, which describes a
truly one dimensional Bose gas.  With regards to band structure,
Fallani {\it et al.}~\cite{Fallani2004}, investigated the dynamic
instability of a BEC in a lattice for various quasimomenta.  By
determining the rate of loss due to heating at a given
quasimomentum, they were able to compare their data with
theoretically predicted instability growth rates.  We emphasize
that our study treats the quasi one dimensional regime in which
the BEC is always in a superfluid state and can be modeled by the
NLS with a periodic potential, as was the case in the experiment
of Fallani {\it et al.}

The case of a BEC trapped in a sinusoidal potential has been
studied theoretically in great detail by a number of
researchers~\cite{Lenz1994,Wu2000,Wu2001,Bronski2001,Bronski2001_2,Diakonov2002,Mueller2002,Wu2002,Deconinck2002,Machholm2003,Massignan2003,Louis2003,Rey2003,Taylor2003,Ostrovskaya2004,Rey2004,Porter2004}.
Analytic results were restricted to certain special
cases~\cite{Bronski2001,Bronski2001_2,Deconinck2002,Porter2004};
the full band structure was obtained numerically, as for example,
by expansion of the wavefunction in Fourier
modes~\cite{Machholm2003}.  Many of these authors considered
stability properties of stationary states.  We will compare our
results to those of Machholm {\it et al.}~\cite{Machholm2003}.  A
generalized periodic potential in the form of a Jacobian elliptic
function has been studied with mathematical
rigor~\cite{Porter2004,Bronski2001,Bronski2001_2}.  A subset of the
solutions to the Kronig-Penney potential has been
found~\cite{Theodorakis1997,Gaididei1997,Taras1999,Alexander2002,Li2004}.
Recently, for example, Li and Smerzi~\cite{Li2004}, investigated
generalized Bloch states for constant phase and zero current.

The article is organized as follows.  In Sec.~\ref{sec:BlochWavefunctions}, the Bloch wave solutions to the stationary NLS with a Kronig-Penney potential are presented.  The quasimomentum-energy bands are detailed for repulsive and attractive condensates in both the weakly and strongly interacting regimes in Sec.~\ref{sec:RepulsiveInteraction} and Sec.~\ref{sec:AttractiveInteraction}.  In Sec.~\ref{sec:DensityProfiles}, the changes in the density profile of the condensate is examined as the quasimomentum changes.  In Sec.~\ref{sec:Limit}, various limiting cases of the NLSE are examined.  The stable and unstable regimes of the bands are studied in detail for both repulsive and attractive condensates in Sec~\ref{sec:Stability}.  Finally, concluding remarks are made in Sec.~\ref{sec:Conclusion}.

\section{The Nonlinear Schr\"odinger Equation and Bloch Waves}
\label{sec:BlochWavefunctions}

In this paper, we consider the mean-field model of a quasi-1D BEC
in the presence of a Kronig-Penney potential, \be V(x)=V_0
\sum_{j=-\infty}^{+\infty}\delta(x- j\,d)\, . \ee where $d$ is the
lattice spacing and $V_0$ is the strength of the potential.  When
the transverse dimensions of the BEC are on the order of its
healing length, and its longitudinal dimension is much longer than
its transverse ones, the 1D NLS~\cite{Carr2000_3} which describes
the stationary states of the mean field of a BEC is given by, \be
-\frac{1}{2}\Psi_{xx} + g|\Psi|^2\Psi + V(x)\Psi = \mu \Psi\, .
\label{eqn:NLS} \ee Note that harmonic oscillator confinement in
the transverse directions with frequency $\omega$ has been assumed
for atoms of mass $m$ and $\mu$ is the eigenvalue, $g$
characterizes the short range pairwise interaction, and $V(x)$ is
an external potential~\cite{Olshanii1998}.  In this paper, a
weakly interacting system is defined by, $gn/V_0 \le 1$, and a
strongly interacting system is defined by, $gn/V_0 \gg 1$.  In
the case where the harmonic oscillator length approaches the
$s$-wave scattering length, $a_s$, the 1D NLS no longer models the
system and a one-dimensional field theory with the appropriate
effective coupling constant must be considered
instead~\cite{Olshanii1998}.  Since $a_s$ is on the scale of
hundreds of angstroms for typical BEC's, this regime is not
relevant to the present study.

In Eq.~(\ref{eqn:NLS}), the length is scaled according to the
lattice spacing, $d$, and energy has been rescaled by $\pi^2/(2
E_0)$, where \be E_0 \equiv \frac{\hbar^2 \pi^2}{2 m d^2}\, , \ee
is the kinetic energy of a particle with wave vector equal to that
at the boundary of the first Brillouin zone.  The variables in
Eq.~(\ref{eqn:NLS}) are defined by, \bea x    &=& \frac{1}{d}x'\,
,
\\
\mu  &=& \frac{\pi^2}{2 E_0} \mu'\, ,
\\
g    &=& \frac{\pi^2}{2 E_0} g'\, ,
\\
V(x) &=& \frac{\pi^2}{2 E_0} V'(x'/d)\, , \eea where the primed
variables contain the physical units of the system.  The
renormalized 1D coupling is, $gn \equiv 2 n a_s \omega_\perp \hbar
m d^2/\hbar^2$, where $n$ is the average density per lattice site.
  Both attractive
and repulsive atomic interactions, {\it i.e.}, $g>0$ and $g<0$,
shall be considered.  The wave function or order parameter $\Psi(x,t)$ has the physical
meaning of, \be \Psi(x,t)=\sqrt{\rho(x,t)}\exp[-i \mu
t+i\phi(x,t)]\, , \label{eqn:GenSol} \ee where $\rho(x,t)$ is the
line density and the superfluid velocity is given by
$v(x,t)=(\hbar/m) \partial \phi(x,t)/\partial x$.

In addition to the NLS, Eq.~(\ref{eqn:NLS}), the normalization of
the wavefunction is given by, \be n=\int_0^1\rho(x')\dd x'\, ,
\label{eqn:number} \ee where $n$ is the number of atoms per
lattice site.  The boundary conditions induced by the
Kronig-Penney potential causes a discontinuity in the derivative
of the wavefunction across each delta function, \be
\partial_x \rho(j+\epsilon)-\partial_x \rho(j-\epsilon)=4 V_0\rho(0)\, ,
\label{eqn:discontinuity}
\ee
where $j$ is an integer and $\epsilon \to 0$.

A brief review is now given of the general solution to
Eq.~(\ref{eqn:NLS}) with no external potential.  We have
previously presented a proof that this represents
the full set of solutions for a constant potential~\cite{Seaman2004}.  Therefore, by using this
complete set of stationary state solutions to the constant
potential case, we can calculate the full set of Bloch solutions
for a lattice.  With the solutions in the form of Bloch waves, the
relationship between the energy per particle and the quasimomentum
of the Bloch waves is determined.

The constant potential solutions are the form of the density in
each lattice site.  The density, $\rho$, and the phase, $\phi$,
are, \bea \rho(x)&=&B+\frac{k^2 b^2}{g} \sn^2(b\,x+x_0,k)\, ,
\label{eqn:density}
\\
\phi(x)&=&\alpha \int_0^x \frac{1}{\rho(x')} \dd x'\, ,
\label{eqn:phase} \eea where $sn$ is a Jacobi elliptic
function~\cite{Abramowitz1964,Milne1950} and the density offset,
$B$, the horizontal scaling, $b$, the translational offset, $x_0$,
and the elliptic parameter, $k$, are free variables.  The Jacobi
elliptic functions are generalized periodic functions
characterized by an additional parameter, $k \in [0,1]$.  In the
limit that $k \to 0$ and $k \to 1$, the Jacobi elliptic functions
become circular and hyperbolic trigonometric functions,
respectively.  The period of the square of the Jacobi elliptic
functions is given by $2 K(k)$, where $K(k)$ is the complete
elliptic integral of the first
kind~\cite{Abramowitz1964,Milne1950}.

Inserting Eqs.~(\ref{eqn:density}) and (\ref{eqn:phase}) into Eqs.~(\ref{eqn:GenSol}) and (\ref{eqn:NLS}), with $V(x)=0$, one finds that the eigenvalue, $\mu$, and phase prefactor, $\alpha$, are given by,
\bea
\mu&=&\frac{1}{2}(b^2(1+k^2)+3 B g)\, ,
\\
\alpha^2&=&B(k^2 b^2/g+B)(b^2+B g)\, . \eea Note that the fact
that $\alpha$ only enters into the equations as $\alpha^2$ implies
that all nontrivial phase solutions, {\it i.e.}, those for which
$\alpha \ne 0$, are doubly degenerate, as $\pm \alpha$ lead to the
same value of the eigenvalue, $\mu$, without otherwise changing
the form of the density or phase.  The boundary conditions are
used to determine the appropriate values of the variables.

After using the methods from the previous paper~\cite{Seaman2004},
the solutions are now represented in the usual Bloch form, \be
\Psi(x)=e^{iqx}f_q(x), \label{eqn:Bloch} \ee where $q$ is the
quasimomentum and $f_q(x)$ has the same period as the lattice,
$f_q(x)=f_q(x+1)$.  By substituting Eq.~(\ref{eqn:Bloch}) into
Eq.~(\ref{eqn:GenSol}), one finds that the density, $\rho$, must
also have the same period as the lattice and that the
quasimomentum and energy per particle can be determined from the
density profile by, \bea q &=& \int_0^1 \frac{\alpha}{\rho(x')}
\dd x'. \label{eqn:quasi}
\\
\frac{E[\Psi]}{n} &=& \frac{1}{n}\int_{0}^{1} \dd x \Big{(}
\frac{1}{2} |\Psi_x|^2 + \frac{g}{2} |\Psi|^4 + V_0
\delta(x)|\Psi|^2 \Big{)}\, . \label{eqn:NLSEnergy} \eea The
quasimomentum is simply the phase jump across each lattice site
and corresponds to the momentum due to the superfluid velocity of
the system.  Since nontrivial phase solutions are degenerate for
the two phase prefactors, $\pm \alpha$, only half of a Brillouin
zone needs to be calculated, {\it i.e.} $0 \le q \le \pi$.  The
second half will then be a reflection around $q=0$.

This paper examines the quasimomentum-energy bands and, therefore,
reduces the problem to a situation where the density is symmetric
around the ends, $x=j$ or $x=j+1$, and the middle, $x=j+0.5$, of
the lattice sites, where $j$ is an integer.  Due to this symmetry,
there are only two possible values for the translational offset,
$x_0$, \be x_0 \in \{-\frac{b}{2}, K(k)-\frac{b}{2}\}\, , \ee
where $K(k)$ is the half period of the density.  The offset forces
the density in the center of each site to be either a minimum or a
maximum of the site, depending on the sign of the interaction.

Since it is computationally intensive to include the integral in
the quasimomentum equation, Eq.~(\ref{eqn:quasi}), with a root
finding algorithm, one of the parameters, $b$, $k$, or $B$, is
varied while the other two are determined from the number
equation, Eq.~(\ref{eqn:number}), and boundary condition,
Eq.~(\ref{eqn:discontinuity}).  The quasimomentum and energy are
evaluated from these parameters and can then be plotted
parametrically.  In the following two sections, we discuss the
energy bands from repulsive and attractive interatomic
interactions.


\section{Repulsive Atomic Interactions}
\label{sec:RepulsiveInteraction}

\begin{figure}[tb]
\begin{center}
\epsfxsize=7.5cm \leavevmode \epsfbox{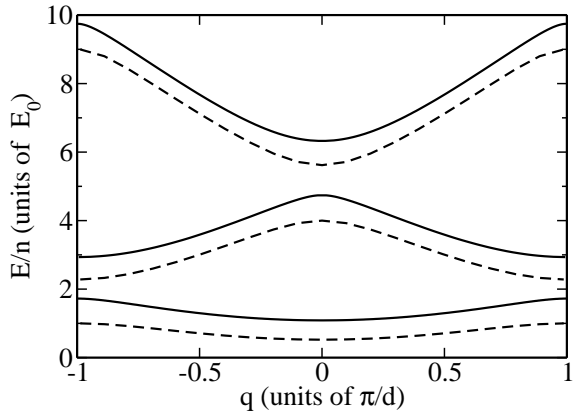}

\caption{Energy per particle as a function of the quasi-momentum for the first three bands of a weakly repulsive condensate ($gn=E_0$) in a repulsive lattice ($V_0=E_0$).  The noninteracting linear band structure is given by the dashed curves.
\label{fig:g1}}
\end{center}
\end{figure}

\begin{figure}[tb]
\begin{center}
\epsfxsize=7.5cm \leavevmode \epsfbox{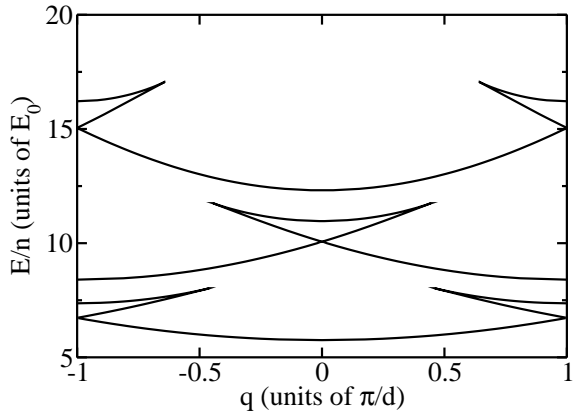}
\caption{Energy per particle as a function of the quasi-momentum for the first three bands of a strongly repulsive condensate ($gn=10 E_0$) in a repulsive lattice ($V_0=E_0$).
\label{fig:g10}}
\end{center}
\end{figure}

The structure of the energy bands is strongly dependent on the
strength and sign of the atom-atom interactions, $g$.  In
Fig.~\ref{fig:g1} and Fig.~\ref{fig:g10}, the energy bands for
specific cases of weak and strong repulsive interactions are
presented, $gn = E_0$ and $gn = 10 E_0$, respectively.  The
condensates are assumed to be in a repulsive lattice, $V_0 = E_0$.
Note that the energies are given as unscaled values so that easy
comparison with previous literature is possible.

%

%

Notice that in Fig.~\ref{fig:g1}, the interaction strength is
small and deviations from the linear band structure are small as
well.  The bands are vertically shifted higher as compared to a
linear system due to the repulsive interactions that increase the
energy of the system.  When the interaction strength is further
increased the band structure becomes quite different.   {\it
Swallowtails}~\cite{Wu2001} appear at the ends of the bands, as in
Fig.~\ref{fig:g10}.  The width of these swallowtails grows as the
interaction strength is increased.  Swallowtails are a general
feature of a nonlinear system in a periodic
potential~\cite{Wu2000} and appear for both repulsive and
attractive interactions.

\begin{figure}[tb]
\begin{center}
\epsfxsize=7.5cm \leavevmode \epsfbox{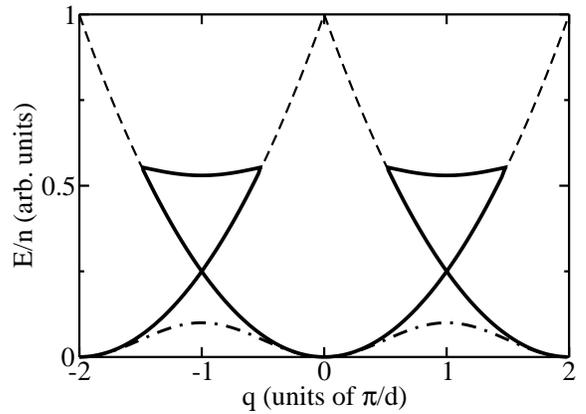}
\caption{Energy per particle as a function of the quasi-momentum wave number for a noninteracting system with no potential (dashed curve), a periodic potential and a small or no interaction (dot-dashed curve) and a periodic potential with a large interaction (solid curve).
\label{fig:example}}
\end{center}
\end{figure}

The presence of these swallowtails is due to the hysteric behavior
of the superfluid condensate.  A thorough discussion of this topic
is given by Mueller~\cite{Mueller2002} (see also references
therein).  For a completely free, noninteracting system, the
energy has a quadratic dependence on the momentum, shown as the
dashed parabolic curves in Fig.~\ref{fig:example}.  Since a
quasimomentum of $2 \pi$ leaves the system unchanged, the
quadratic dependence is repeated centered around integer multiples
of $2\pi$.  When a periodic potential is added to the system,
bands in the energy are formed, shown as the dot-dash sinusoidal
curve in Fig.~\ref{fig:example}.  These bands may be found by
solving the linear Schr\"odinger equation.  An interacting
condensate, however, is a superfluid and can therefore screen out
the periodic potential~\cite{Taylor2003}.  The energy band will
then appear similar to that of a free particle, shown as a solid
swallowtail curve in Fig.~\ref{fig:example}, until a critical
point.  Since there is a critical velocity, determined by the
condensate sound speed, the energy band must terminate.  If this
velocity allows the quasimomentum to pass the edge of the
Brillouin zone, there are then two separate energy minima.  This
demands that there be a saddle point separating them and hence the
three stationary states.  For a Kronig-Penney potential, just as
for a sinusoidal potential, when a swallowtail appears there are
two spinodal points, where the number of energy extrema changes,
at the edges of the swallowtail.  These are points where the local
minimum and local maximum energy converge to the same energy.  As the
interaction strength is increased, the width of the swallowtail
can increase without limit.  Eventually, the swallowtail will
become large enough to cross the higher bands.  In this case,
there is a degeneracy between bands.

%

For interacting systems in periodic potentials, there is a minimum
interaction strength for which the swallowtails in the energy
bands can exist~\cite{Mueller2002}.  This can, in general, be
dependent on both the strength of the potential as well as the
band that is being discussed.  For an optical lattice it was shown
that the onset of the swallowtail for the lowest band of a
repulsive condensate occurs when the interaction strength and the
potential are equal.  For higher bands the relationship no longer
becomes analytic~\cite{Machholm2003}.  For the Kronig-Penney
lattice potential, the critical value for the onset of the
swallowtails is not dependent on the band under consideration or
the sign of the interaction.  Numerically, we are able to
determine that the onset occurs when $gn = 2 V_0$.  The factor of
two is not present for the optical lattice potential.  This holds
true in all situations except for the lowest band of an attractive
condensate, as will be explained in the following section.

The energy bands are slightly different when the condensate is in
an attractive potential, i.e. $V_0<0$.  At the Brillouin zone
boundary, the energy gap between bands is proportional to $V_0$
for a weakly interacting system.  The density is an equally spaced
array of solitons.  With a quasimomentum of $q=\pi$ in the lowest
band, the condensate has one dark soliton\cite{Kivshar1998} per
lattice site with nodes at the delta functions.  The potential,
therefore, does not contribute to the total energy of the system
since the delta function potential only depends on the density at
that position.  The second band with quasimomentum of $q=\pi$,
however, has a density that is offset by half a
period.  There is then a dark soliton in the center of each
lattice site.  The effect of the potential lowers the energy by
$\rho(0) V_0$.  The second band then can cross the first band for
a condensate in an attractive potential.  In contrast, the bands
are separated by an energy $\rho(0) V_0$ for a repulsive
potential.  As the interaction strength increases, the effects of
the potential become less noticeable and the bands are no longer
degenerate.  Note that repulsive and attractive sinusoidal
potentials create the same band structure.  The difference in the current system arises
since there are two length scales associated with a Kronig-Penney
potential, the lattice spacing and delta function width.

\section{Attractive Atomic Interactions}
\label{sec:AttractiveInteraction}

The energy bands for an attractive condensate in a repulsive
potential with a small interaction strength have a qualitatively
similar form as for a weakly repulsive condensate.  Note that the
attractive bands are, however, lower in energy than the repulsive
bands.  The attractive bands are pushed down from the linear case
due to the attractive interaction strength.  A strongly attractive
condensate, however, has several qualitative differences compared
to a strongly repulsive condensate.

\begin{figure}[tb]
\begin{center}
\epsfxsize=7.5cm \leavevmode \epsfbox{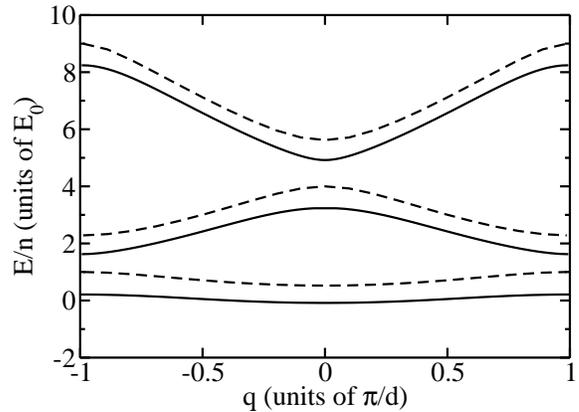}
\caption{Energy per particle as a function of the quasi-momentum wave number for the first three bands of a weakly attractive condensate ($gn=-E_0$) in a repulsive lattice ($V_0=E_0$).  The noninteracting linear band structure is given by the dashed curves.
\label{fig:g_1}}
\end{center}
\end{figure}

In Fig.~\ref{fig:g_1}, the band structure for a small attractive
interaction, $gn=-E_0$, in a repulsive potential, $V_0 = E_0$, is
illustrated.  The band structure for this system is almost
identical to that of the weakly repulsive condensate,
Fig.~\ref{fig:g1}.  This is to be expected since a weak
interaction should only cause small perturbations to a
noninteracting system.

%

\begin{figure}[tb]
\begin{center}
\epsfxsize=7.5cm \leavevmode \epsfbox{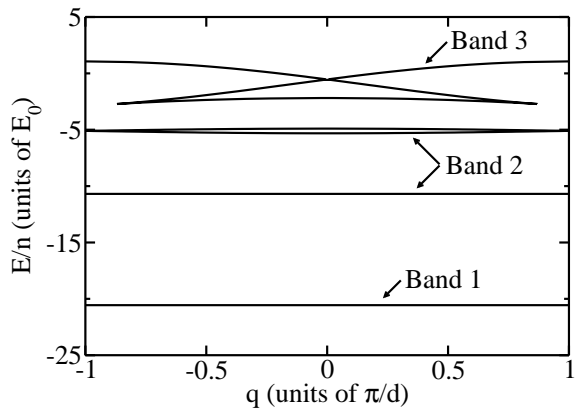}
\caption{Energy per particle as a function of the quasi-momentum wave number for the first three bands of a strongly attractive condensate ($gn=-10 E_0$) in a repulsive lattice ($V_0=E_0$).  The first band does not have a swallowtail.  However, the second and third bands do have swallowtails.  The swallowtail in the second band appears as a loop with an unattached curve since an attractive condensate has a maximum width for the swallowtails, which in the case of the second band is a width of $\pi$.
\label{fig:g_10}}
\end{center}
\end{figure}

In Fig.~\ref{fig:g_10}, the band structure for a large attractive
interaction, $gn=-10 E_0$, in a repulsive potential, $V_0 = E_0$,
is illustrated.  The energy band for this system is vastly
different from the strongly repulsive case.  The swallowtails in
the bands are now on the upper band at the band gaps as opposed to
the lower bands at the band gaps as they were for the repulsive
case in Fig.~\ref{fig:g10}.  The first energy band never has a
swallowtail.  This is because the swallowtail must be on the lower
portion of the band and below the center of the band there is no
quadratic energy dependence for the swallowtail to follow, see
Fig.~\ref{fig:example}.  Therefore, the lowest energy band only
has a swallowtail for a condensate with repulsive interactions.

%

The second band of a strongly interacting attractive condensate
looks quite different than that of a strongly interacting
repulsive condensate.  For an attractive condensate, after the
initial critical value of the interaction strength is reached, a
swallowtail in the band starts to form, as in the third band of
Fig.~\ref{fig:g_10}.  As the interaction strength increases, the
width of the swallowtail also increases.  Eventually, another
critical value is reached where the width of the swallowtail in
$q$ reaches $\pi$ and runs into the band edge.  If the interaction
strength is further increased, the width of the swallowtail
increases but the quasimomentum that should be less than zero
becomes imaginary due to the form of $\alpha$.  This represents a
nonphysical solution and is contrary to the assumption that the
phase is real in Eq.~(\ref{eqn:GenSol}).  Therefore, part of the
band is absent and the band appears as two separate pieces, a loop
and a separate line.  These are both marked as Band 2 in
Fig.~\ref{fig:g_10}.  The band takes on this appearance since the
swallowtail is attempting to extend lower than the minimum in the
adjacent quadratic energy dependence.  In a similar fashion to the
first band, the solutions are not physical.  The portion of the
swallowtail that is before this edge is physical, however,
and so a third energy extremum is still present.  The third band
will exhibit the same behavior when the width of the swallowtail
reaches $2\pi$.  In general, the $n^{th}$ band will appear similar
when the swallowtail reaches a width of $(n-1)\pi $ since the
swallowtail will then have reached the minimum of the
corresponding free particle energy.  It should be noted that this
phenomenon does not occur for a repulsive interaction since there
is no extremum to limit the growth of the swallowtail.

The energy bands are slightly different when the condensate is in
an attractive potential.  In a manner similar to a repulsive condensate in an attractive potential, for a sufficiently weak interaction strength and large potential strength, the first and second bands can become degenerate.

It should be noted that for a single attractive delta function on the infinite line, there is a critical repulsive interaction above which the impurity can no longer bind the condensate~\cite{Seaman2004}.  However, for a periodic system of delta functions, there is no longer a critical value.  This can best be understood by considering a condensate of a finite size bound by a single delta function.  As the repulsive interaction increases, the condensate spreads out until it reaches the next delta function.  This continues indefinitely as the interaction strength increases.  There is, therefore, no transition from a bound to unbound state as there was with a single impurity.  The symmetry breaking nature of a single impurity allows for the transition that is not present with the symmetric lattice.
\section{Density Profiles}
\label{sec:DensityProfiles}

As density is the primary experimental observable for BEC's, the
change in the density profile as the quasimomentum is varied is
important.  The case of weak interactions creates similar changes
in the density profile and energy band structure independent of
the sign of the interaction.  This is because the interaction
energies are of the same magnitude as the potential.  For strong
interactions, the density is in general more sharply peaked for an
attractive condensate and flatter for a repulsive condensate.
However, the general way in which the density changes is
qualitatively similar.

\begin{figure}[tb]
\begin{center}
$\begin{array}{c}
\epsfxsize=7.5cm  \epsffile{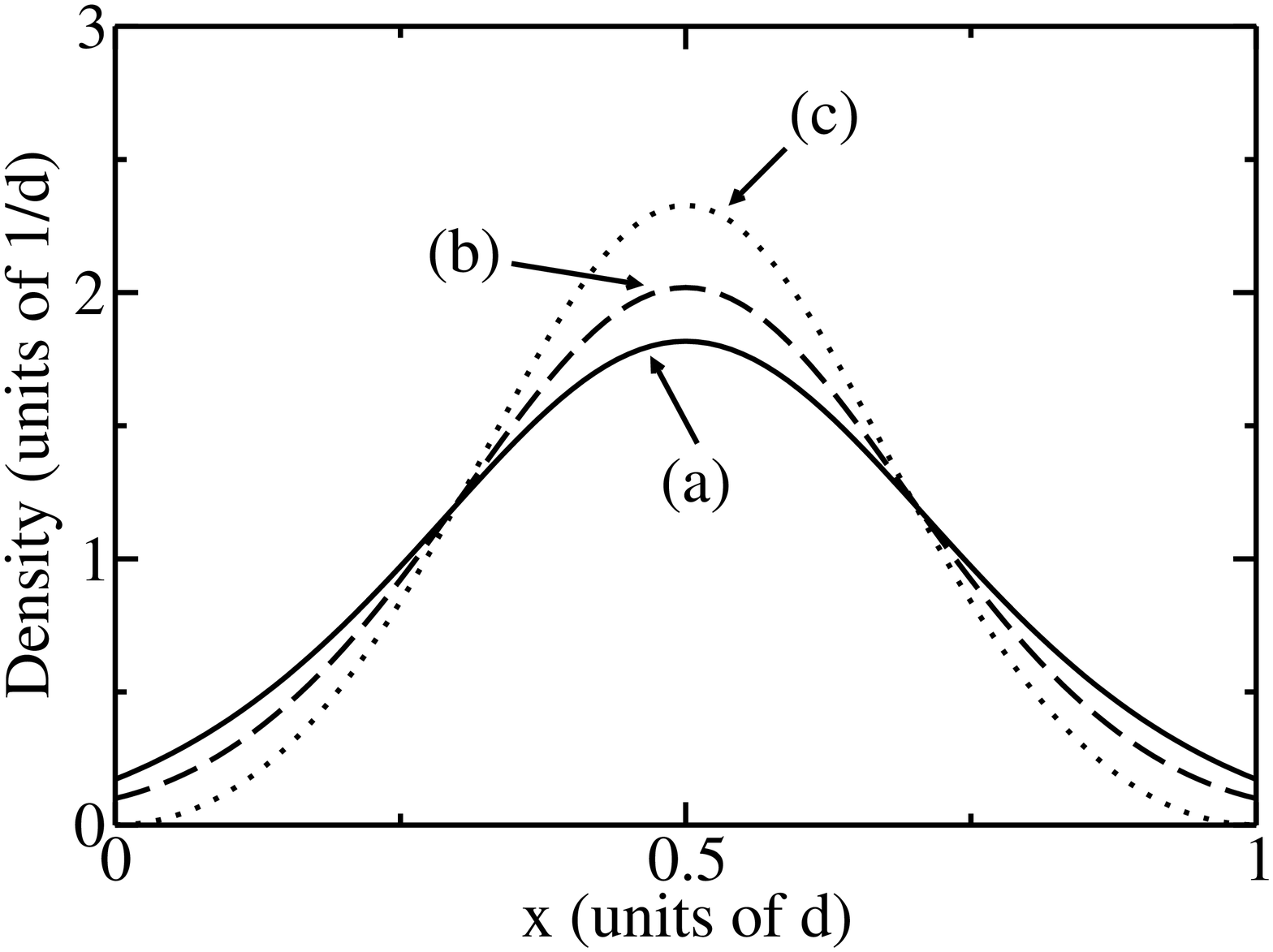} \\ \\ \\
\epsfxsize=7.5cm  \epsffile{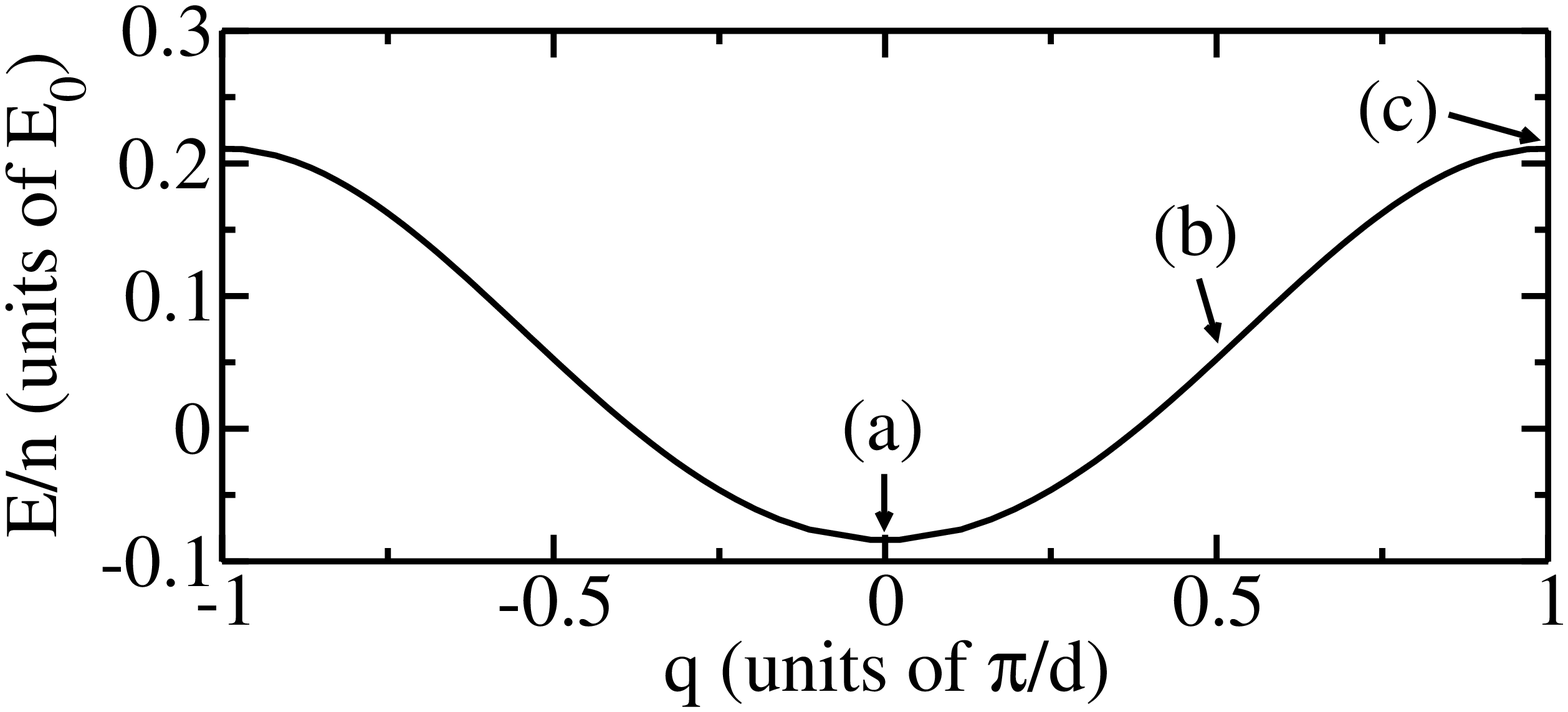}
\end{array}$
\caption{Changes in the density of a weakly attractive, $gn=-E_0$,
condensate associated with different positions on the first band.
The solid curve represents the density when $q=0$.  The dashed
curve represents the density when $q=\pi/2$.  The dotted curve
represents the density  when $q=\pi$. The lower plot shows the
corresponding positions on the first energy band.
\label{fig:density_g-1}}
\end{center}
\end{figure}

The density changes in the first band of the weakly attractive
condensate are shown for three different quasimomenta in
Fig.~\ref{fig:density_g-1}.  The solid curve, dashed curve, and
dotted curve represent the density profile for $q=0$, $q=\pi/2$,
and $q=\pi$, respectively.  To understand the energy bands in
terms of the density profile, the three terms of the energy in
Eq.~(\ref{eqn:NLSEnergy}) should be discussed.  The kinetic,
interaction and potential energy per particle are given by, \bea
\frac{E_{k}}{n} &=& \frac{1}{n}\int_{0}^{1} \frac{1}{2}
|\Psi_x|^2 \dd x\, ,
\\
\frac{E_{i}}{n} &=& \frac{1}{n}\int_{0}^{1}\frac{g}{2} |\Psi|^4 \dd x\, ,
\\
\frac{E_{p}}{n} &=& \frac{1}{n}\int_{0}^{1} V_0
\delta(x)|\Psi|^2 \dd x\, , \eea respectively.  Notice that the density
at the origin monotonically decreases as the quasimomentum is
increased.  Therefore the potential energy will also monotonically
decrease due to the delta function at $x=j$, where $j$ is an
integer.  Since the condensate is attractive, the interaction
energy decreases monotonically as the quasimomentum is increased
since the density is becoming more peaked.  The kinetic energy
monotonically increases as the quasimomentum increases since the
variations in the density become larger.  For a weakly repulsive
condensate, the density will also become more strongly peaked as
the quasimomentum is increased and will thus have an interaction
energy that increases.  The larger variations in the density will
then increase the kinetic energy as well.

%

\begin{figure}[tb]
\begin{center}
$\begin{array}{c}
\epsfxsize=7.5cm \epsffile{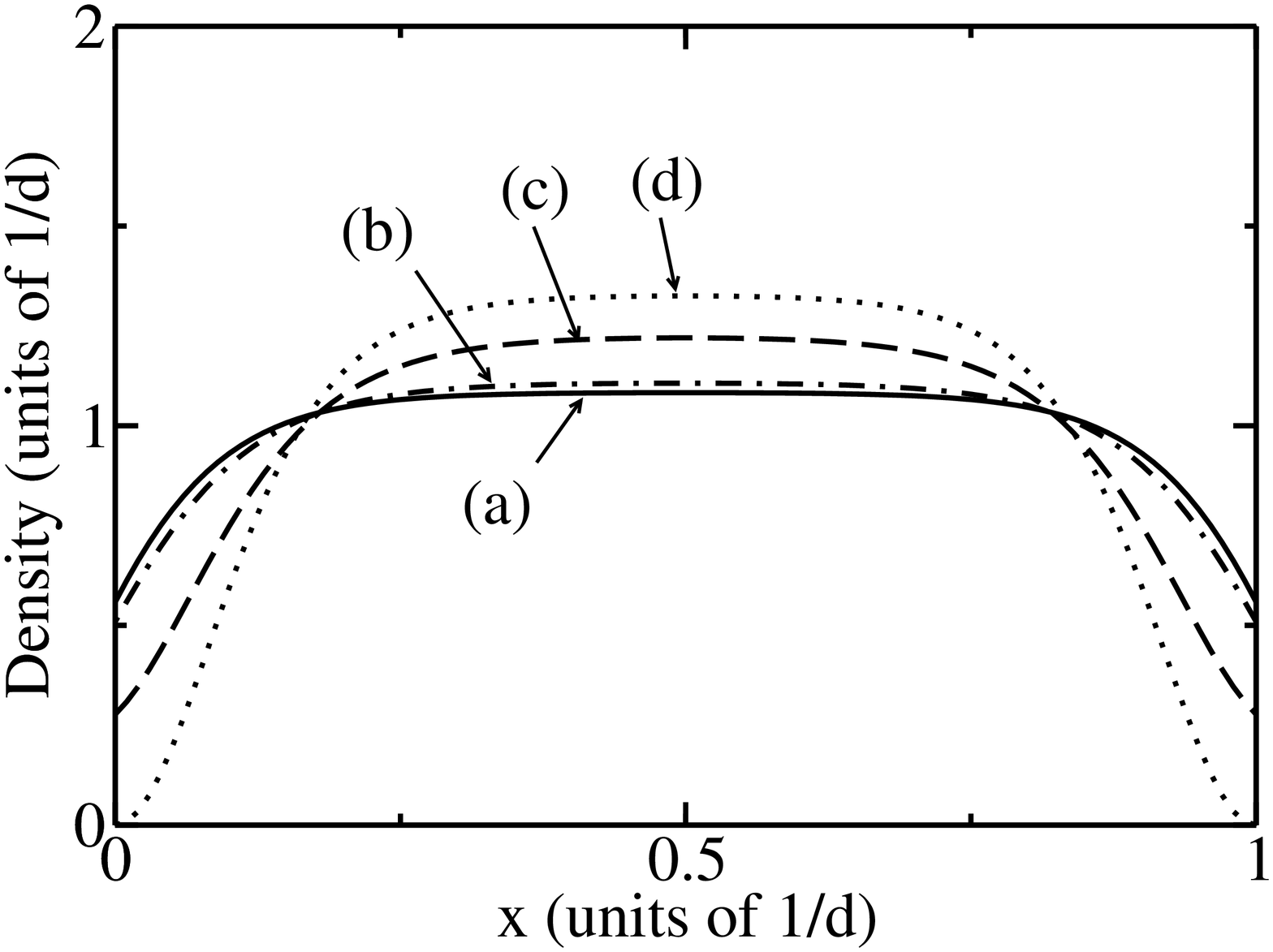}\\ \\ \\
\epsfxsize=7.5cm \epsffile{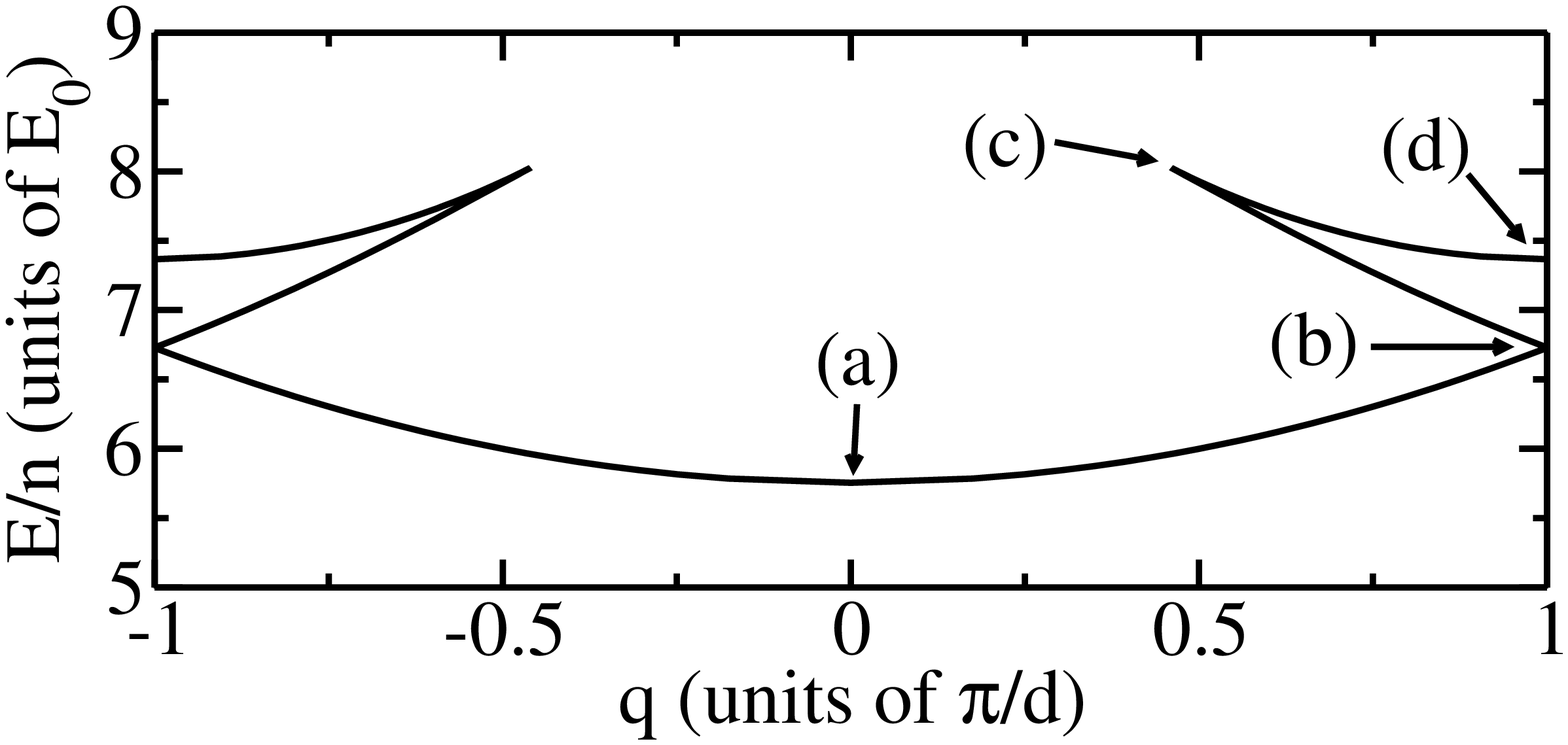}
\end{array}$
\caption{Changes in the density of a strongly repulsive
condensate, $gn=10 E_0$, associated with different positions on
the first band.  The  solid curve represents the density when
$q=0$.  The  dot-dashed curve represents the density when $q=\pi$
at the bottom of the swallowtail.  The  dashed curve represents
the density when $q=0.46 \pi$ at the end of the swallowtail.  The
dotted curve represents the density when $q=\pi$ at the top of the
swallowtail.  The lower plot shows the corresponding positions on
the first energy band. \label{fig:density_g_10}}
\end{center}
\end{figure}

The case of the density variations associated with the first band
of a strongly interacting repulsive condensate is shown in
Fig.~\ref{fig:density_g_10}.  The change in the density is
qualitatively similar to that of the weakly attractive case.  The
solid line represents the density profile for $q=0$.  The dot-dash
curve represents the density when $q=\pi$ at the bottom of the
swallowtail.  The dashed curve represents the density when $q=0.46
\pi$ at the end of the swallowtail.  The dotted curve represents
the density when $q=\pi$ at the top of the swallowtail.  The
density at the origin again monotonically decreases as the
quasimomentum increases and therefore the potential energy also
increases monotonically.  The interaction energy decreases
monotonically as the band is traversed since the density becomes
more uniform.  The kinetic energy follows the same
qualitative path as the total energy.  It is therefore the kinetic
energy that has the greatest influence on the energy bands.

%

In the case of strongly attractive interactions, the lowest band
consists of well separated bright solitons centered in each
lattice site.  As the quasimomentum changes from zero to $\pi$,
the density changes very little.  This is easily noted in the
energy spectrum of the lowest band which changes on the order of a
tenth of a percent~\ref{fig:g_10}.

\section{Limiting Cases}
\label{sec:Limit}

It is now shown how the solutions to the NLS connect
to the solutions of the linear Schr\"odinger equation as well as
to the discrete nonlinear Schr\"odinger equation.  The linear
Schr\"odinger equation is important since it describes electron
motion through crystals.  The discrete nonlinear Schr\"odinger
equation describes superfluid systems where the potential strength
is much greater than the interaction strength.  It is a limit of
the Bose-Hubbard model when the Hamiltonian is projected onto a
coherent state~\cite{Rey2004}.

\subsection{Linear Schr\"odinger Equation Limit}
\label{subsec:LinearLimit}

The energy dispersion relation of the noninteracting regime can be
determined from the solutions of the interacting system.  In the
linear limit, the interaction strength, $g$, and elliptic
parameter, $k$, must vanish such that, \be \lim_{g,k \to 0}
\frac{k^2 b^2}{g}=A\, , \ee where $A$ is the height of the density
fluctuations.  The elliptic parameter must vanish since in this
limit the Jacobi elliptic functions become circular trigonometric
functions, which is the appropriate linear form of the density.
The linear solutions are then of the form, \be \rho(x) = B + A
\sin^2(b x + j \pi/2-b/2)\, , \ee where $j$ is a positive integer
and $b$ is the horizontal scaling.  The offset $j\pi/2$ can be
dropped, since the effect in the square of the sine function is to
change it into either the square of a sine or a square of a cosine
and this merely alters the values of $A$ and $B$. The
quasimomentum equation, Eq.~(\ref{eqn:quasi}), can be integrated
exactly in the linear limit to find \be \cos^2(q/2) =
\frac{1}{1+(1+r)\tan^2(b/2)}\, , \label{eqn:linearQuasi} \ee
where $r\equiv A/B$ is the ratio of the density modulation
coefficient, $A$, to the constant density offset, $B$.  The
equation that governs the discontinuity in the derivative of the
density at the delta functions, the boundary conditions given by
Eq.~(\ref{eqn:discontinuity}), can be solved for the ratio $r$,
\be r=-\frac{V_0}{b \sin(b/2)\cos(b/2)+V_0\sin^2(b/2)}\, .
\ee The quasimomentum can then be related to the horizontal
scaling, $b$, through the relation, \be \cos(q) = \cos(b)+\frac{V_0}{b}\sin(b)\, . \ee This dispersion relationship
coincides with that of the linear Schr\"odinger equation in
introductory quantum mechanics texts~\cite{Liboff2003}.  At the
edges of the Brillouin zone, $q= 0$ and $q= \pi$, the heights
of the bands and the gaps can be extracted.

\subsection{Discrete Nonlinear Schr\"odinger Equation Limit}
\label{subsec:DiscreteLimit}

It is now shown how the solutions to the NLS connect
to the solutions of the discrete nonlinear Schr\"odinger equation
(DNLS).  The DNLS describes a system where the density is
localized in predominantly non-overlapping regions described by
the Wannier functions $w_j$, \be w_j(x-x_j) \equiv
\frac{1}{\sqrt{N}}\int e^{iqx_j}\Psi(x) \dd q\, , \ee where $x_j$
is the position of the $j^{th}$ lattice site, $N$ is the total
number of lattice sites, and the integral is over the first
Brillouin zone.  Since the regions are non-overlapping, it is
required that, \be \int w_j(x) w_k(x) \dd x \approx \delta_{jk}\,
, \ee where $w_j(x)$ is the Wannier function localized at site $j$
and $\delta_{jk}$ is the Kronecker delta
function~\cite{Christodoulides2003}.  This is also known as the
tight-binding approximation~\cite{Smerzi2003}.  In addition, it is
assumed that particles can only hop to a nearest neighbor site,
{\it i.e.} particles in the $j^{th}$ site can only move to the $(j
\pm 1)^{th}$ site.  Therefore, all integrals of the form, \be \int
(\partial_x w^*_j)(\partial_x w_k) \dd x\, , \ee vanish unless $k
\in \{j,j \pm 1\}$.  One possible way to create a system like this
is to make the strength of the potential sufficiently strong in
comparision to the interaction strength of the condensate, $V_0/g
n \gg 1$.

The DNLS may be extracted from the NLS by using a single band
approximation and writing the wavefunction as a linear combination
of site-localized Wannier functions, \be \Psi(x,t) = \sum_j
\psi_j(t) w_j(x)\, , \label{eqn:Wannier} \ee where $\psi_j(t)$ is
the probability amplitude to find a particle at site $j$ and
$w_j(x)$ is the density profile that is localized on site $j$.
When Eq.~(\ref{eqn:Wannier}) is substituted into
Eq.~(\ref{eqn:NLS}), and the spatial degrees of freedom are
integrated over, the evolution of the system becomes governed by,
\be i \partial_t \psi_j = \epsilon_j \psi_j + J_j
(\psi_{j+1}+\psi_{j-1}) + U_j |\psi_j|^2\psi_j\, , \ee where
$\epsilon_j$ is the on-site energy of the $j^{th}$ site, $J$ is
the hopping strength, and $U$ is the interaction strength, with
each defined by, \bea \epsilon_j &\equiv& V_0
|w_j(0)|^2+\frac{1}{2}\int|\partial_x w_j|^2 \dd x\, ,
\\
J_j &\equiv& \frac{1}{2} \int (\partial_x w^*_j)(\partial_x w_{j \pm 1}) \dd x\, ,
\\
U_j &\equiv& g \int |w_j|^4 \dd x\, . \eea The stationary states
of the DNLS are given by plane waves, $\psi_j = \psi_0 e^{i(p
j-\mu t)}$, where $p$ is the momentum and the eigenvalue, $\mu$,
is given by, \be \mu = U |\psi_0|^2+\epsilon + 2 J \cos(p)\, . \ee
The subscript $j$ has been dropped since all Wannier functions
have the same profile, only translationally shifted to the
appropriate site.  The energy of the condensate for each site is
given by, \be E_{site} = \frac{1}{2} U |\psi_0|^4 + |\psi_0|^2
\epsilon + 2 |\psi_0|^2 J \cos(p)\, .
\label{eqn:WannierSiteEnergy} \ee If the Wannier functions are
normalized to unity, then $\psi_0 = \sqrt{n}$ and the equation for
the energy per particle becomes, \be \frac{E_{site}}{n} =
\frac{1}{2} UN + \epsilon + 2 J \cos(p)\, .
\label{eqn:WannierParticleEnergy} \ee The Wannier functions for
the Kronig-Penney potential can be extracted from
Eq.~(\ref{eqn:density}) by taking the limit when the potential
strength approaches infinity.

It should be noted that a large attractive interaction in a
repulsive potential will have a similar effect as a large
repulsive potential and hence the DNLS can be used to describe the
system.  In this case, the large attractive interaction causes the
density of each site to approach a hyperbolic secant squared
profile, i.e., a bright soliton.  Hence, $k \to 1$ and $B \to
-b^2/g$.  The normalization of the density fixes $b=gn/2$ and the
quasimomentum becomes zero.  This density profile is the
equivalent of the Wannier functions discussed earlier.  The
on-site energy is calculated to be, \be \epsilon=- gn V_0 \exp[g
n /2] + \frac{g^2 n^2}{24}\, . \ee The interaction term becomes,
\be U = -\frac{g^2 n^2}{6}\, . \ee With the assumption that $|g|$
is large, the hopping term is given by, \be J = \frac{g^2 n^2}{8}
(4 + gn) \exp[gn/2]\, . \ee The DNLS energy,
Eq.~(\ref{eqn:WannierSiteEnergy}), becomes consistent with the NLS
energy, Eq.~(\ref{eqn:NLSEnergy}), to within three tenths of a
percent for the first band of the strongly attractive case, $gn =
-10 E_0$ and $V_0 = E_0$.  When the on-site energy, interaction
energy, and hopping strength are all assumed to remain constant,
the maximum error in energy remains within three tenths of a
percent across the entire band.  The energy is slightly
underestimated at zero quasimomentum and is slightly overestimated
at $\pi$ quasimomentum.  In this instance, knowing the density
profile at one value of the quasimomentum allows for extension to
the full spectrum of quasimomentum.  When the interaction energy
becomes strong enough that only terms to highest order of $g$ need
to be considered, the total energy becomes $-g^2 n^2/24$ and the
variation across the band becomes $-g^3 n^3/2 \exp[gn/2]$, which
is essentially flat.

\section{Stability}
\label{sec:Stability}

We proceed to study the stability of the Bloch states and to
determine the stable regions of the bands are determined.  In
addition to stable solutions, solutions that have instability
times much longer than experimental time scales can be observed in
experiments.  Recent studies of the stability of condensates in a
periodic potential have focused on linear energetic and dynamic
stability, also called Landau
stability~\cite{Taras1999,Bronski2001,Machholm2004,Machholm2003,Fallani2004}.
In contrast, we consider the full response of the condensate to
stochastic perturbations.  In order to numerically simulate the
NLS with a periodic potential, a ring geometry is used with a
quantized phase.  To ensure that the phase quantization does not
effect the stability properties, enough lattice sites were used to
allow for many rotations of the phase such that $N q = 2\pi j$, where $q$ is the quasimomentum, $N$ is
the number of sites nad $j$ is an interger.  The outcome of the stability analysis is
independent of the number of sites for sufficiently large number
of sites.  In most cases, $j=4$ was found to be adequate to
extract the correct stability properties.

The delta functions are simulated by single point distortions in
the potential grid.  They are also implemented by using boxes of
different widths with their area normalized to create the
appropriate potential strength.  The size of the boxes did not
influence the stability properties until the width became
approximately $10\%$ of the size of the healing length, $\xi
\equiv \hbar/\sqrt{2gn}$.  The NLS is evolved using a variable
step fourth-order Runge-Kutta algorithm in time and a filtered
pseudo-spectral method in space.  The noise introduced into the
simulations comes from the round off error associated with
numerical simulations.  To ensure that the form of the noise from
round off error did not affect the stability properties of the
system, initial stochastic white noise of various levels was
introduced into the Fourier spectrum.  For levels significantly
greater than the round off noise, the stability times approached
those from the round off noise.  Introduction of white noise at
the level in the eighth significant digit produced the same
instability times as the round off noise, which effects the
sixteenth significant digit.  All simulations were performed over
time scales longer than experimental lifetimes of the BEC, which
are on the order of seconds.

The time at which the onset of instability occurs is determined by
the effective variance in the Fourier spectrum, \be \sigma(t)
\equiv \sqrt{\frac{\sum(f(p,t)-f(p,0))^2}{2 \sum(f(p,0))^2}}\, ,
\ee where $f(p,t)$ is the Fourier component of the wavefunction at
momentum $p$ and time $t$ and the sum is over the momentum grid.
This quantity determines how different the Fourier spectrum is
compared to the original stationary state.  It vanishes when the
two spectrums are identical and approaches unity when there are no
Fourier components in common.  When $\sigma(t)$ reaches $0.5$,
{\it i.e.} $50\%$ of the Fourier spectrum is different than the
original, the system is considered to have become unstable.

Unless otherwise noted, for the stability analysis the lattice
spacing is given by $d=1$ $\mu$m, the length scale with which
current optical lattices are created.  In addition, all
instability time scales will be given for $^{87}$Rb.

\subsection{Attractive Atomic Interactions}
\label{subsec:AttractiveInteraction}

With an attractive interaction of $gn=-E_0$ in a repulsive
potential of $V_0 = E_0$ (see Fig.~\ref{fig:g_1}), the lowest
energy solution, zero quasimomentum in the lowest band, has a
lifetime greater than experimental time scales.  However, when
even a slight harmonic perturbation to the potential is added to
the initial time step, for instance a harmonic frequency of $120$
Hz which is approximately the experimental trapping
frequency~\cite{Moritz2003} in the longitudinal dimension, the
condensate becomes unstable on time scales on the order of $1.5$
ms.  This is short compared to the lifetime of a
BEC~\cite{Moritz2003}, but is still observable.

When the quasimomentum of the first band is increased, the
stability of the system becomes dependent on the effective mass.
The effective mass, $m^*$, is defined by\cite{SmerziComment},
\be
m^* \equiv\frac{1}{\partial^2 E/\partial q^2}\, .
\ee 
The effective mass is the mass that the particle appears to have if the potential was
not being considered~\cite{Altmann1970}. 
The sign of the effective mass can be transferred to the interaction strength,
changing an attractive interaction to a repulsive interaction.
Therefore, when the quasimomentum increases and the energy band
becomes concave down, $m^*<0$, the system enters a regime of
stability.  The system remains stable even in the presence of the
harmonic perturbation.

For zero quasimomentum in the second band, the system immediately
possesses periodic variations in the phase and density.  There is
an additional underlying instability that occurs on the order of
$5$ ms that destroys the periodicity of the system.  It would be
expected that this part of the band be stable since there is a
negative effective mass but the oscillations due to the two bright
solitons per lattice site force the system to be unstable.  The
oscillations, although periodic in time, create a larger
underlying instability to grow.

The stability properties of a strongly attractive condensate are similar to those of a weakly attractive condensate.  The stability of the first band is determined by the effective mass while higher bands always go unstable.  Therefore, for an attractive condensate, the system of Bloch waves is stable only if there is one soliton per lattice site and the effective mass is negative.

\subsection{Repulsive Atomic Interactions}
\label{subsec:RepulsiveInteraction}

\begin{figure}[tb]
\begin{center}
\epsfxsize=9.5cm \leavevmode \epsfbox{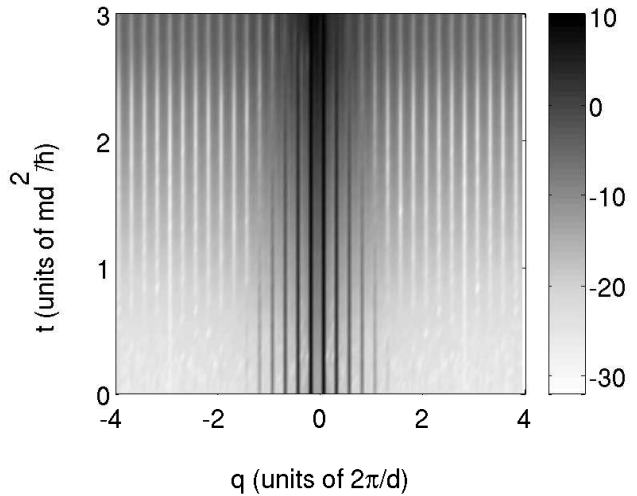}
\caption{Logarithm of the Fourier spectrum during the time
evolution.  Time is in units of $\hbar/E_0$ and distance in units
of the lattice spacing. \label{fig:instabilityM}}
\end{center}
\end{figure}

\begin{figure}[tb]
\begin{center}
\epsfxsize=7.5cm \leavevmode \epsfbox{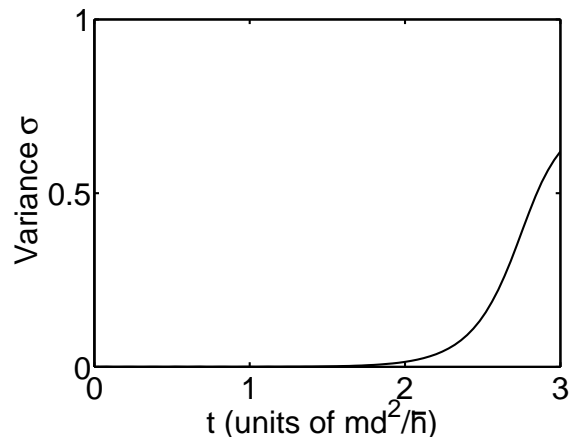} \caption{The
time evolution of the effective variance of the momentum density,
$\sigma$.  Time is in units of $\hbar/E_0$. \label{fig:variance}}
\end{center}
\end{figure}

Similar to an attractive condensate, a repulsive condensate only
has stable regions on the first band.  For a weakly
interacting repulsive condensate, $gn= E_0$ and $V_0 = E_0$, the
effective mass in the first band is positive between $q=0$ and
$q=\pi/2$.  The effective mass then becomes negative for larger
quasimomentum since the energy becomes concave down and hence the
system becomes unstable.  For a quasimomentum of $q=9 \pi/16$ the
instability time is $10$ ms and reduces to $2$ ms for $q=\pi$.  In
this regime, with negative effective mass, the ground state is an
envelope soliton that can spread over many lattice sites.  These
types of states are called gap
solitons~\cite{Lenz1994,Eiermann2004,Hilligsoe2002,Louis2003} and
only occur in interacting systems.  Figure~\ref{fig:instabilityM}
presents the unstable evolution of the weakly repulsive condensate
 in the first band with a quasimomentum of $q=\pi$ in Fourier space.
  Notice that the
instabilities arise from perturbations around the primary Fourier
components of the wavefunction.  In Fig.~\ref{fig:variance}, the
effective variance, $\sigma$, is plotted as a function of
evolution time and the system becomes unstable around $2$ ms.  The
second band becomes unstable when $q=0$ in $0.500$ ms and in $6$
ms for $q=\pi$.  Therefore, the system is stable in the first band
with positive effective mass, $m^*>0$, and unstable elsewhere.  This is consistent with the effects that the effective mass has on stability in systems described by the lowest band DNLS.  In
a work by Fallani {\it et al.}~\cite{Fallani2004}, the instability
time of a condensate in a lattice was measured by using an
RF-shield to remove the hottest atoms produced by the heating
created in the sample due to instability.  The loss rate, given by
the inverse of the lifetime, should then be qualitatively similar
to the instability time.  Our calculations are consistent with
these experimentally observed loss rates of a BEC in an optical
lattice~\cite{Fallani2004}.

%

%

Due to the presence of the swallowtails, the strongly interacting
system provides different stability regimes.  For a repulsive
condensate, $gn = 10 E_0$, in a repulsive lattice, $V_0 = E_0$,
the main section of the first band, as well as the lower portion
of the swallowtail, have positive effective mass and remain
stable.  The upper portion of the swallowtail, as discussed in
Sec.~\ref{sec:RepulsiveInteraction}, is an energy maximum and is
not expected to remain stable.  In fact, the instability time of
the upper portion of the swallowtail is approximately $0.2$ ms,
independent of the actual quasimomentum.

\begin{figure}[tb]
\begin{center}
\epsfxsize=7.5cm \leavevmode \epsfbox{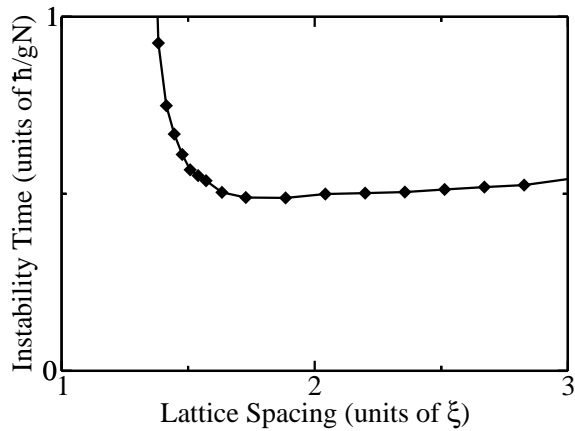} \caption{The time to instability as the lattice spacing is varied.
Time is in unit of $\hbar/gn$ and distance in units of the healing
length, $\xi = \hbar/\sqrt{2 gn}$. \label{fig:timescales}}
\end{center}
\end{figure}

The size of the lattice spacing can influence the time scales for
which the system becomes unstable.  In Fig.~\ref{fig:timescales},
the time to instability is presented as a function of the lattice
spacing for a repulsive condensate, $gn=E_a$, in a repulsive
lattice, $V_0 = E_a$, with a quasimomentum of $q=\pi$, where $E_a
\equiv \hbar^2\pi^2/2m(1\mu m)^2$.

%

There is a minimum instability time as the lattice spacing varies
that occurs when the lattice spacing is approximately twice the healing length,
$\xi$.  The time to instability is given by half of the
interaction strength time, $gn/\hbar$.  When the lattice spacing
is much larger than the healing length, the density becomes
extremely flat except at the delta functions, where dark solitons
form.  Since the lattice spacing is large, the dark solitons are
far apart and are effectively noninteracting pinned solitons. Dark
solitons are known to be robustly
stable~\cite{Kivshar1998,Bronski2001_2} and, therefore, for a
lattice spacing much larger than the healing length the system
should become stable.  For lattice spacing smaller than twice the
healing length, the condensate has difficulty distinguishing
between the separate delta functions and sees closer to a constant
potential.  In this regime, the kinetic energy becomes much
greater than the interaction energy and potential energy since
variations in the density occur on the length scale of $d/2$,
which is less than the healing length.  The system then becomes
effectively free and noninteracting and, therefore, approaches
stability.

\section{Discussion and Conclusions}
\label{sec:Conclusion}

The full set of stationary states of a Bose-Einstein condensate in
a Kronig-Penney lattice potential, with period commensurate with
the lattice, have been presented analytically in the form of Bloch
waves for both repulsive and attractive interactions.  The
quasimomentum energy bands were found to exhibit a cusp at the
critical interaction strength $g n = 2 V_0$, where $g$ is the
interatomic interaction strength, $n$ is the linear density, and
$V_0$ is the lattice potential strength.  For larger interaction
strengths, swallowtails form in the bands.  These swallowtails
have the same qualitative form as for a sinusoidal potentail and
exhibit the same stability properties.

Both attractive and repulsive condensates were found to be
dynamically stable only in the first band and when the effective
interaction, sgn$(m^*)g$, was positive.  Also, even in the first
band, the upper edges of swallowtails were always unstable.
It should be noted that the effective mass does not influence the stability of the higher bands since they are always unstable.  
Therefore, for an attractive condensate, the only stable Bloch
states exist in the first band between the quasimomentum where
$m^*$ becomes negative and $q=\pi$.  A repulsive condensate is
only stable in the first band from $q=0$ to the quasimomentum
where $m^*$ becomes negative.  Higher bands will always be
unstable, for both attractive and repulsive condensates.  When
solutions became unstable, our numerically studies consistently
observed that the instabilities originated around the primary
Fourier components of the wavefunction.  This is in agreement with
the formal proof of the instability of constant phase Bloch-type
solutions by Bronski {\it et al.}~\cite{Bronski2001_2}.  The
instability time was found to be a function of the lattice
constant.  If the delta functions are spaced either smaller than
or much larger than the healing length of the condensate, the
solutions had instability times longer than lifetime of the BEC.
Thus experiments could access formally unstable sections of the
energy bands, and, by controlling the ratio of the healing length
to the lattice constant, directly observe the dynamics of
instability.  The results of our stability analysis are consistent
with the experimental work performed by Fallani {\it et
al.}~\cite{Fallani2004}, in which the loss rate, the inverse of
the lifetime, was determined by removing the hottest atoms with an
RF-shield.

An interesting phenomenon to note is that for a repulsive
condensate when a swallowtail is present, the entire energy band
is concave up and, hence, the effective mass is always positive.
This is in contrast to a weakly repulsive condensate, when the
concavity of the energy band changes, creating a region of
negative effective mass.  Therefore, there is a maximum
interaction strength for which gap
solitons~\cite{Lenz1994,Hilligsoe2002,Louis2003,Christodoulides2003,Eiermann2004}
can be formed since they require a negative effective mass, which
does not occur when a swallowtail is present.  The maximum
interaction energy is given by $gn = 2 V_0$, the strength at
which swallowtails appear.

Stationary solutions to the NLS with a Kronig-Penney potential
need not take the form of Bloch states.  Solutions with a period
which is an integer multiple of the lattice period have been shown
to exist for the sinusoidal potential~\cite{Machholm2004} and are
expected also to be present for the Kronig-Penney potential.
Envelope solutions, such as gap solitons, also play an important
role in other systems modelled by the NLS and have been observed
in BEC's~\cite{Eiermann2004}.  The analytic methods which we have
described here are equally applicable to these solution types and
will form the subject of future study~\cite{Seaman2005}.

\begin{center}
{\bf Acknowledgments}
\end{center}

We acknowledge helpful discussions with John Cooper, Chris Pethick, Ana Maria Rey, Augusto Smerzi, and Eugene Zaremba.  Support is acknowledged for
B.T.S. from the National Science Foundation and for L.D.C. and
M.J.H. from the U.S. Department of Energy, Office of Basic Energy
Sciences via the Chemical Sciences, Geosciences and Biosciences
Division.


\bibliographystyle{prsty}

\begin{thebibliography}{10}

\bibitem{Altmann1970}
S.~L. Altmann, {\em Band Theory of Metals: The Elements} (Pergamon PRess,
  Oxford, 1970).

\bibitem{Usmanov2004}
R.~A. Usmanov and L.~B. Ioffe, Phys. Rev. B {\bf 69},  214513  (2004).

\bibitem{Christodoulides2003}
D.~N. Christodoulides, F. Lederer, and Y. Silberberg, Nature {\bf 424},  817
  (2003).

\bibitem{Greiner2001}
M. Greiner, I. Bloch, O. Mandel, T.~W. Hansch, and T. Esslinger, Appl. Phys. B
  {\bf 73},  769  (2001).

\bibitem{Fallani2003}
L. Fallani, F.~S. Cataliotti, J. Catani, C. Fort, M. Modugno, M. Zawada, and M.
  Inguscio, Phys. Rev. Lett. {\bf 91},  240405  (2003).

\bibitem{Fallani2004}
L. Fallani, L. De~Sarlo, J.~E. Lye, M. Modugno, R. Saers, C. Fort, and M.
  Inguscio, Phys. Rev. Lett. {\bf 93},  140406  (2004).

\bibitem{Eiermann2004}
B. Eiermann, T. Anker, M. Albiez, M. Taglieber, P. Treutlein, K.~P. Marzlin,
  and M.~K. Oberthaler, Phys. Rev. Lett. {\bf 92},  230401  (2004).

\bibitem{Eiermann2003}
B. Eiermann, P. Treutlein, T. Anker, M. Albiez, M. Taglieber, K.-P. Marzlin,
  and M.~K. Oberthaler, Phys. Rev. Lett. {\bf 91},  060402  (2003).

\bibitem{Esslinger2003}
T. Esslinger and K. Molmer, Phys. Rev. Lett. {\bf 90},  160406  (2003).

\bibitem{Greiner2002}
M. Greiner, O. Mandel, T. Esslinger, T.~W. Hansch, and I. Bloch, Nature {\bf
  415},  39  (2002).

\bibitem{Anderson1998}
B.~P. Anderson and M.~A. Kasevich, Science {\bf 282},  1686  (1998).

\bibitem{Hagley1999}
E.~W. Hagley, L. Deng, M. Kozuma, J. Wen, K. Helmerson, S.~L. Rolston, and
  W.~D. Phillips, Science {\bf 283},  1706  (1999).

\bibitem{Ovchinnikov1999}
Y.~B. Ovchinnikov, J.~H. M\"uller, M.~R. Doery, E.~J.~D. Vredenbregt, K. Helmerson, S.~L.
  Rolston, and W.~D. Phillips, Phys. Rev. Lett. {\bf 83},  284  (1999).

\bibitem{Roberts1998}
J.~L. Roberts, N.~R. Claussen, J.~P. Burke, Jr., C.~H. Greene, E.~A. Cornell,
  and C.~E. Wieman, Phys. Rev. Lett. {\bf 81},  5109  (1998).

\bibitem{Inouye1998}
A. Inouye, M.~R. Andrews, J. Stenger, H.-J. Miesner, D.~M. Stamper-Kurn, and W.
  Ketterle, Nature {\bf 392},  151  (1998).

\bibitem{Gross1961}
E.~P. Gross, Nuovo Cimento {\bf 20},  454  (1961).

\bibitem{Pitaevskii1961}
L.~P. Pitaevskii, Zh. Eksp. Teor. Fiz. {\bf 40},  646  (1961).

\bibitem{Seaman2004}
B.~T. Seaman, L.~D. Carr, and M.~J. Holland, cond-mat/0410345  (2004).

\bibitem{Astrakharchik2004}
G.~E. Astrakharchik and L.~P. Pitaevskii, Phys. Rev. A {\bf 70},  013608
  (2004).

\bibitem{Radouani2004}
A. Radouani, Phys. Rev. A {\bf 70},  013602  (2004).

\bibitem{Bogdan1997}
M.~M. Bogdan, A.~S. Kovalev, and I.~V. Gerasimchuk, Low Temp. Phys. {\bf 23},
  145  (1997).

\bibitem{Machholm2003}
M. Machholm, C.~J. Pethick, and H. Smith, Phys. Rev. A {\bf 67},  053613
  (2003).

\bibitem{Wu2002}
B. Wu, R.~B. Diener, and Q. Niu, Phys. Rev. A {\bf 65},  025601  (2002).

\bibitem{Mueller2002}
E.~J. Mueller, Phys. Rev. A {\bf 66},  063603  (2002).

\bibitem{Rey2003}
A.~M. Rey, K. Burnett, R. Roth, M. Edwards, C.~J. Williams, and C.~W. Clark, J.
  Phys. B {\bf 36},  825  (2003).

\bibitem{Anderson1995}
M.~H. Anderson, J.~R. Ensher, M.~R. Matthews, C.~E. Wieman, and E.~A. Cornell,
  Science {\bf 269},  198  (1995).

\bibitem{Davis1995}
K.~B. Davis, M.-O. Mewes, M.~R. Andrews, N.~J. van Druten, D.~S. Durfee, D.~M.
  Kurn, and W. Ketterle, Phys. Rev. Lett. {\bf 75},  3969  (1995).

\bibitem{Meystre1999}
P. Meystre and M. Sargent~III, {\em Elements of Quantum Optics}, 3rd ed.
  (Springer, Berlin, 1999).

\bibitem{Cirac2004}
J.~I. Cirac and P. Zoller, Phys. Today {\bf 57},  38  (2004).

\bibitem{Girardeau1960}
M. Girardeau, J. Math. Phys. {\bf 1},  516  (1960).

\bibitem{Paredes2004}
B. Paredes, A. Widera, V. Murg, O. Mandel, S. Folling, Cirac, G.~V.
  Shlyapnikov, T.~W. H\"ansch, and I. Bloch, Nature {\bf 429},  277  (2004).

\bibitem{Kinoshita2004}
T. Kinoshita, T. Wenger, and D.~S. Weiss, Science {\bf 305},  1125  (2004).

\bibitem{Lenz1994}
G. Lenz, P. Meystre, and E.~M. Wright, Phys. Rev. A {\bf 50},  1681  (1994).

\bibitem{Diakonov2002}
D. Diakonov, L.~M. Jensen, C.~J. Pethick, and H. Smith, Phys. Rev. A {\bf 66}, 013604 (2002).

\bibitem{Massignan2003}
P. Massignan and M. Modugno, Phys. Rev. A {\bf 67},  023614  (2003).

\bibitem{Ostrovskaya2004}
E.~A. Ostrovskaya and Y.~S. Kivshar, Phys. Rev. Lett. {\bf 92},  180405
  (2004).

\bibitem{Louis2003}
P.~J.~Y. Louis, E.~A. Ostrovskaya, C.~M. Savage, and Y.~S. Kivshar, Phys.
  Rev. A {\bf 67},  013602  (2003).

\bibitem{Wu2001}
B. Wu and Q. Niu, Phys. Rev. A {\bf 64},  061603(R)  (2001).

\bibitem{Rey2004}
A.~M. Rey, Ph.D. thesis, University of Maryland at College Park, 2004.

\bibitem{Deconinck2002}
B. Deconinck, B.~A. Frigyik, and J.~N. Kutz, J. Nonlinear Sci. {\bf 12},  169
  (2002).

\bibitem{Taylor2003}
E. Taylor and E. Zaremba, Phys. Rev. A {\bf 68},  053611  (2003).

\bibitem{Wu2000}
B. Wu and Q. Niu, Phys. Rev. A {\bf 61},  023402  (2001).

\bibitem{Bronski2001}
J.~C. Bronski, L.~D. Carr, B. Deconinck, and J.~N. Kutz, Phys. Rev. Lett. {\bf
  86},  1402  (2001).

\bibitem{Bronski2001_2}
J.~C. Bronski, L.~D. Carr, B. Deconinck, J.~N. Kutz, and K. Promislow, Phys.
  Rev. E {\bf 63},  036612  (2001).

\bibitem{Porter2004}
M.~A. Porter, P.~G. Kevrekidis, and B.~A. Malomed, Physica D {\bf 196},  106
  (2004).

\bibitem{Li2004}
W.-D. Li and A. Smerzi, Phys. Rev. E {\bf 70},  016605  (2004).

\bibitem{Taras1999}
D. Taras-Semchuk and J.~M.~F. Gunn, Phys. Rev. B {\bf 60},  13139  (1999).

\bibitem{Alexander2002}
S.~A. Alexander and R.~L. Coldwell, Int. J. of Qu. Chem. {\bf 86},  325
  (2002).

\bibitem{Theodorakis1997}
S. Theodorakis and E. Leontidis, J. Phys. A {\bf 30},  4835  (1997).

\bibitem{Gaididei1997}
Y.~B. Gaididei, P.~L. Christiansen, K.~O. Rasmussen, and M. Johansson, Phys.
  Rev. B {\bf 55},  R13365  (1997).

\bibitem{Carr2000_3}
L.~D. Carr, M.~A. Leung, and W.~P. Reinhardt, J. Phys. B {\bf 33},  3983
  (2000).

\bibitem{Olshanii1998}
M. Olshanii, Phys. Rev. Lett. {\bf 81},  938  (1998).

\bibitem{Abramowitz1964}
{\em Handbook of Mathematical Functions}, edited by M. Abramowitz and I.~A.
  Stegun (National Bureau of Standards, Washington, D.C., 1964).

\bibitem{Milne1950}
L.~M. Milne-Thomson, {\em Jacobian Elliptic Function Tables} (Dover
  Publications Inc., New York, 1950).

\bibitem{Kivshar1998}
Y.~S. Kivshar and B. Luther-Davies, Physics Reports {\bf 298},  81  (1998).

\bibitem{Liboff2003}
R.~L. Liboff, {\em Introductory Quantum Mechanics}, fourth ed. (Addison Wesley,
  San Francisco, 2003).

\bibitem{Smerzi2003}
A. Smerzi and A. Trombettoni, Phys. Rev. A {\bf 68},  023613  (2003).

\bibitem{Machholm2004}
M. Machholm, A. Nicolin, C.~J. Pethick, and H. Smith, Phys. Rev. A {\bf 69},
  043604  (2004).

\bibitem{Moritz2003}
H. Moritz, T. St\"oferle, M. K\"ohl, and T. Esslinger, Phys. Rev. Lett. {\bf
  91},  250402  (2003).

\bibitem{SmerziComment}
Note one can also define an effective mass associated with the chemical
  potential bands as in Li and Smerzi~\cite{Li2004}.

\bibitem{Hilligsoe2002}
K.~M. Hilligsoe, M.~K. Oberthaler, and K.-P. Marzlin, Phys. Rev. A {\bf 66},
  063605  (2002).

\bibitem{Seaman2005}
B.~T. Seaman, L.~D. Carr, and M.~J. Holland, work in progress  (2005).


\end{thebibliography}

\end{document}